\documentclass[twocolumn,pre,a4paper,superscriptaddress,showpacs,showkey,nofootinbib]{revtex4}
\usepackage{amsfonts,amsmath,amssymb,graphicx,epsfig}

\def\chemin{./}

\newcommand{\pic}[3]{\includegraphics[clip=true,width=#2 \linewidth,angle=#3]{\chemin #1}}

\newcommand{\figcap}[2]{\caption{\label{#1}\em #2}}

\newcommand{\bmini}[1]{\begin{minipage}{#1 \linewidth}}
\newcommand{\emini}{\end{minipage}}

\begin{document}
\title{Growth activity during fingering in a porous Hele Shaw cell}
\author{Grunde L\o voll}
\affiliation{Department of Physics, University of Oslo, Norway}
\affiliation{Department of Physics, NTNU Trondheim, Norway}
\affiliation{Laboratoire de G\'eologie, \'Ecole Normale Sup\'erieure, Paris, France}
\author{Yves M{\'e}heust}
\affiliation{Department of Physics, NTNU Trondheim, Norway}
\affiliation{Laboratoire de G\'eologie, \'Ecole Normale Sup\'erieure, Paris, France}
\affiliation{Department of Physics, University of Oslo, Norway}
\author{Renaud Toussaint}
\affiliation{Department of Physics, University of Oslo, Norway}
\affiliation{Department of Physics, NTNU Trondheim, Norway}
\author{Jean Schmittbuhl} \affiliation{Laboratoire de
G\'eologie, \'Ecole Normale Sup\'erieure, Paris, France}
\author{Knut J\o rgen M\aa l\o y} \affiliation{Department of Physics, University of
Oslo, Norway} 

\date{\today}

\begin{abstract}
 We present in this paper an experimental study of the invasion
 activity during unstable drainage in a 2D random porous medium, when
 the (wetting) displaced fluid has a high viscosity with respect to
 that of the (non-wetting) displacing fluid, and for a range of almost
 two decades in capillary numbers corresponding to the transition
 between capillary and viscous fingering. We show that the invasion
 process takes place in an active zone within a characteristic
 screening length $\lambda$ from the tip of the most advanced
 finger. The invasion probability density is found to only depend on
 the distance $z$ to the latter tip, and to be independent of the
 value for the capillary number $C_a$.  The mass density along the
 flow direction is related analytically to the invasion probability
 density, and the scaling with respect to the capillary number is
 consistent with a power law. Other quantities characteristic of the
 displacement process, such as the speed of the most advanced finger
 tip or the characteristic finger width, are also consistent with
 power laws of the capillary number.  The link between the growth
 probability and the pressure field is studied analytically and an
 expression for the pressure in the defending fluid along the cluster is derived.
 The measured pressure are then compared with the
 corresponding simulated pressure field using this expression for the
 boundary condition on the cluster.
\end{abstract}



\maketitle

%
%
\section{Introduction}
Different types of unstable fluid displacements in porous media play
an important role in many natural and commercial
processes~\cite{Bear72,Dullien92}. Developments of a better
understanding of these processes therefore has a broad scientific
interest as well as potentially huge economical benefits. The complex
patterns observed in such processes have been extensively studied and
modeled over the last decades, see
\cite{Bear72,Dullien92,Sahimi93,Sahimi95,Lenormand88,Lenormand89,Saffman58}
and references therein.
 
The geometry of the displacement structures observed in immiscible two
phase flow are in general controlled by the competition between
viscous forces, gravitational forces, capillary forces; those various
forces act
on scales ranging from the pore scale to the system size. The relative
wettabilities, viscosities, and densities of the fluids, as well as
the heterogeneity of the underlying porous media, play an important
role in the competition process. The relative magnitudes of viscous
and capillary forces (on pore scale) are quantified through the
dimensionless {\em capillary number} $C_a =(\mu_w \, v_f
\,a^2)/(\gamma \, \kappa)$ where $\mu_w$ is the viscosity in the
wetting (displaced) fluid, $v_f$ is the filtration speed, $a$ is the
characteristic pore size, $\gamma$ is the interface tension, and
$\kappa$ is the permeability of the porous medium.

In this paper we address a drainage experiment in which non-wetting
air displaces a high viscous wetting glycerin/water solution in a
horizontal two-dimensional porous medium; hence, gravity has no
influence on the displacement. We investigate the crossover regime
between the regime of slow displacement for which capillary forces
control the dynamics of the invasion process and the geometry of the
resulting invasion structure (capillary fingering), and that of fast
displacements where viscous forces are dominant (viscous fingering).
We emphasize on the dependence of the invasion probability density
$\phi$, or activity, on the distance to the most advanced finger tip
along the interface. The invasion probability density $\phi$ is the
growth probability of the invasion structure; it is fundamental
because both the structure and the dynamics are controlled by this
function. Growth probability has been discussed extensively in the
past for DLA (Diffusion Limited Aggregation)
simulations~\cite{Witten81,Plischke84,Plischke85,Meakin86b,Halsey86,Amitrano86}
where it was found to be a multifractal distribution or a harmonic
measure~\cite{Meakin86b,Halsey86,Amitrano86,Mandelbrot74}.
A strong analogy exists between the structures obtained by DLA and by
viscous fingering in a porous medium, as was first pointed out by
Paterson~\cite{Paterson84}. However, the dynamics of the two
processes differ in that there is no surface tension for the DLA, in contrast 
to drainage in porous media where surface tension gives
rise to capillary pressure thresholds at the pore scale. The
capillary pressure threshold values, introduce a lower cut off for the
invasion probabilities, even for fast flows. In the slow displacement
limit for which $v_f \simeq 0$, the invasion process is entirely
controlled by the fluctuations of the capillary threshold distribution
inside the porous medium ~\cite{Lenormand85,Lenormand89}.

Imbibition experiments (wetting fluid displacing a non wetting fluid)
were previously performed in a quasi two-dimensional system
~\cite{Stokes86,Weitz87}, where the width of the viscous fingers was
measured to scale with the capillary number as
~\cite{Stokes86,Weitz87} $w_f \propto C_a^{-0.5}$. This scaling
relation was explained by a strong dynamic component of the capillary
pressure~\cite{Weitz87}. We do not observe a strong dynamic component
of the capillary pressure in our experiments (see below). The geometry
of the invader for drainage is also significantly different from the
invader structure of imbibition
~\cite{Stokes86,Weitz87,Lenormand83,Lenormand90}.

In this study we study experimentally the growth probability density
$\phi(z)$ as a function of the distance $z$ (in the flow direction) from
the most advanced finger tip, and its dependence on the extraction
speed (or capillary number). We also investigate experimentally the
mass density $n(z)$ along the flow direction of the invader, and
confront the behavior of the measured $\phi(z)$ and $n(z)$ to what we
expect from analytical arguments. A calculation of the z dependence of
the pressure on the surface of the invader is presented, which
yields the z dependence of the capillary pressure and shows a direct link
with the measured growth probability density. Pressure measurements
are performed in the model and compared with pressures simulated 
by solving the Laplace equation with this pressure boundary
condition. Other features characteristic of the displacement, such as
outermost tip velocity and the width of the invasion fingers, are also
investigated.

The present article is organized as follows. We first present the
experimental method (section~\ref{sec:exp_method}). We then discuss the experimental
results (section~\ref{sec:results}), before concluding (section~\ref{sec:discussion}).

%
%
\section{Experimental method}
\label{sec:exp_method}

The experimental setup is shown in Fig.~\ref{fig_model}. The porous
model consists of a mono-layer of glass beads of diameter $a = 1$~mm
which is randomly spread between two contact papers
\cite{Maloy85,Meheust2002}. The model is a transparent rectangular box
of dimensions $L \cdot W$ and thickness $a$.

Two models of widths $W=430$~mm and $W=215$~mm have been used in the
experiments; their other characteristics were identical. They are
respectively referred to in the rest of the article as the ``wide''
and the ``narrow'' model.

To prevent bending of the model a $2$~cm thick glass plate and a
$2$~cm thick Plexiglas plate are placed on top of the model. To squeeze
the beads and the contact paper together with the upper plate, a mylar
membrane mounted on a $2.5$~cm thick Plexiglas plate, below the model,
is kept under a $3.5$~m water pressure as a ``pressure cushion''. The
upper and the lower plates are kept together by clamps, and the side
boundaries are sealed by a rectangular silicon rubber packing. Milled
inlet and outlet channels are made in the upper Plexiglas plate. The
distance between the inlet and outlet channels define the length of
the model $L=840$~mm. One should also note that a few beads are
removed from a small region near the center of the inlet channel, to
initiate the invader in the center of the inlet. This is done to avoid
edge effects appearing when the invader grows to the lateral
boundaries of the model. The porosity of the models is measured to be
$0.63$ and the permeability is $\kappa=(0.0166 \pm 0.0017) \cdot
10^{-3}$ cm$^2=(1685 \pm 175)$ Darcy.

The defending wetting fluid used in all our experiments is a
$90\%-10\%$ by weight {\em glycerol-water} solution dyed with $0.1\%$
Negrosine to increase the contrast between the colored fluid and the
invader. {\em Air} is used as the invading non-wetting fluid. The {\em
wetting} glycerol-water solution has a viscosity of $\mu_w \approx
0.165$ Pa.s and a density of $\rho_w = 1235$ kg.m$^{-3}$ at room
temperature. The corresponding parameters for the {\em non-wetting}
air are $\mu_{nw} = 1.9\cdot 10^{-5}$ Pa.s and $\rho_{nw} = 116$
kg.m$^{-3}$. The viscous ratio is thus $M = \mu_{nw}/\mu_{w} \sim
10^{-4}$. The surface tension between these two liquids is $\gamma =
6.4 \cdot 10^{-2}$ N.m$^{-1}$. The temperature in the defending fluid
is controlled and measured at the outlet of the model during each
experiment, so as to accurately estimate the viscosity of the wetting
fluid.

The absolute pressure in the wetting liquid is measured in the outlet
channel and at a point at a distance of $280$~mm (in the flow direction)
from the inlet channel and $38$~mm from the left boundary (looking in
the flow direction) using {\em Honeywell 26PCAFlow-Through} pressure
sensors.

\begin{figure}
 \pic{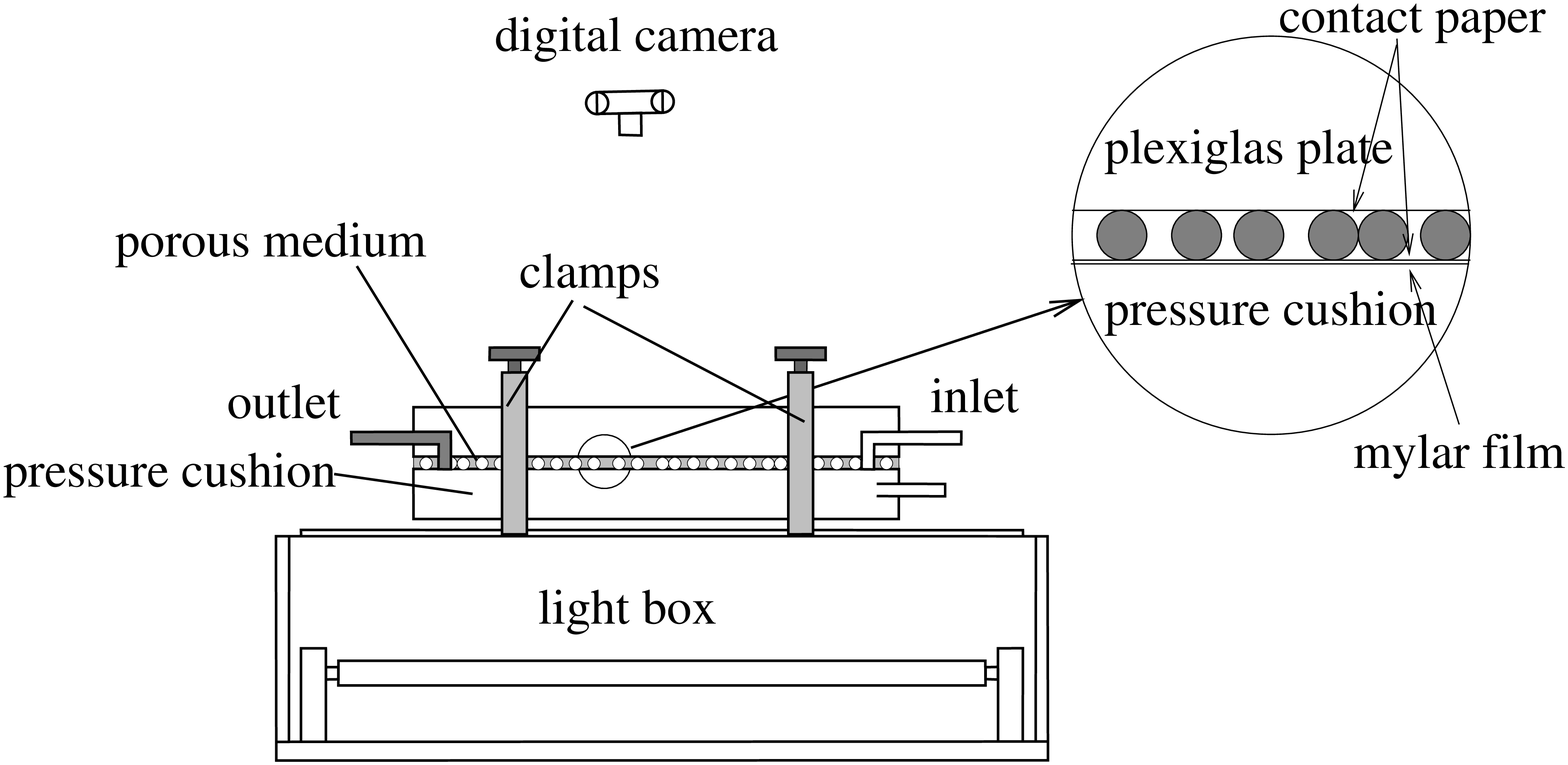}{1}{0} \figcap{fig_model}{Sketch of the
 experimental setup with the light box for illumination, the porous
 model and the digital camera. The porous medium is sandwiched between
 two contact papers and kept together with a ``pressure cushion''.}
\end{figure}

\begin{figure*}
        \pic{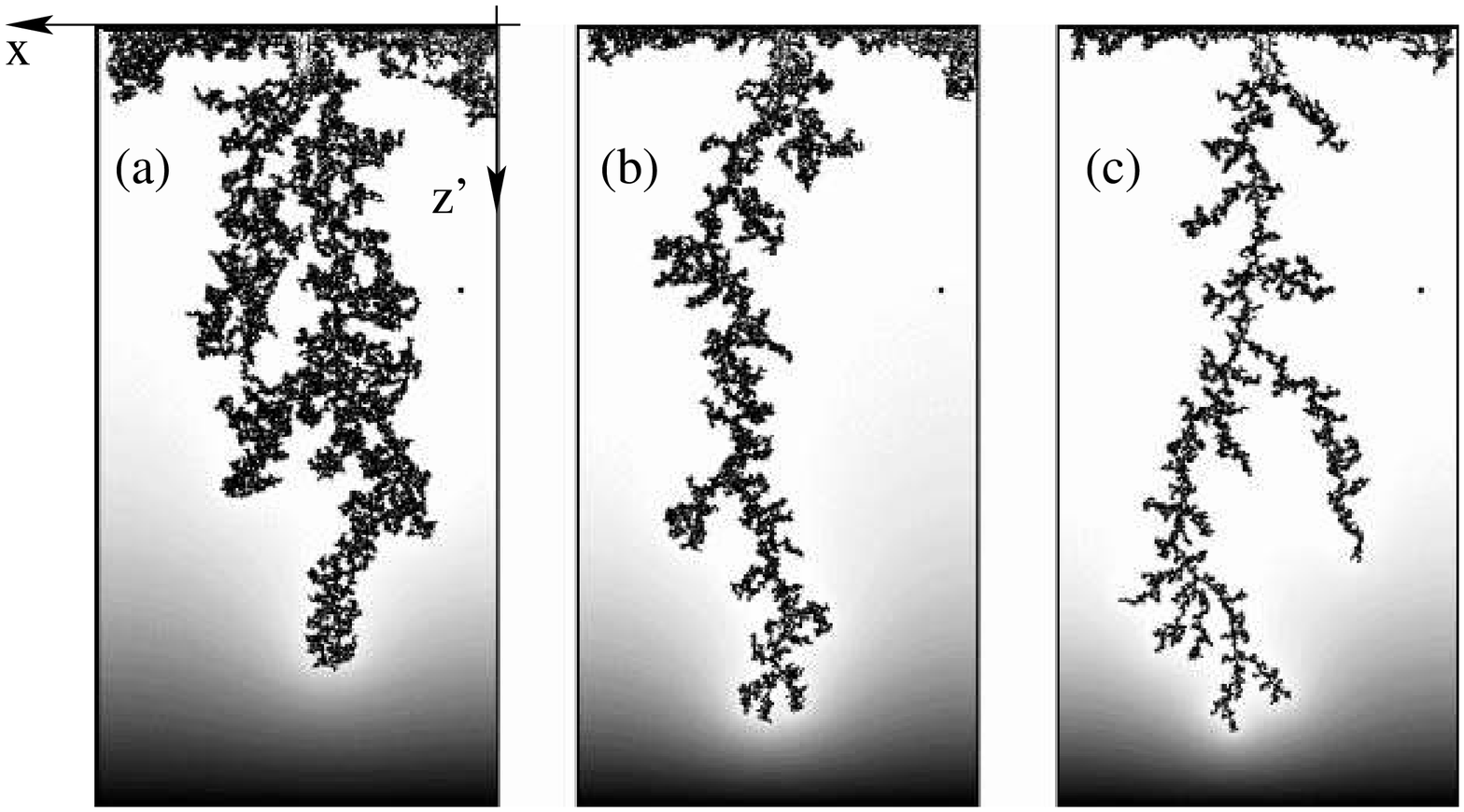}{0.95}{0}
 \figcap{fig_observations}{
 Displacement structures obtained for
 different withdrawal rates: {\bf (a)} $C_a = 0.027$, 
 {\bf (b)} $C_a = 0.059$ 
 and {\bf(c)} $C_a = 0.22$. 
 The images have been treated
 to separate the two phases, excluding the invasion structure close to
 the inlet. The black frame denotes the outer boundaries of the model,
 while the black spot close to the right edge of the model denotes the
 position of the pressure sensor. The simulated pressure field is
 shown superimposed on the image.  Dark shadings correspond to low
 pressures while light shadings correspond to high pressures.
}
\end{figure*}

The invader is visualized by illuminating the model from below with a
light box and taken pictures with a {\em Kodak DCS 420 CCD} camera,
which is controlled by a computer over a SCSI connection. This
computer records both the pictures and the pressure measurements.
Each image contain $1536 \times 1024$ pixels, which corresponds to a
spatial resolution of $0.55$ mm per pixel or $\sim 3.22$ pixels per
pore ($1$ mm$^2$); the color scale contains $256$ Gray levels. The
Gray level distribution of the image presents two peaks corresponding
respectively to the white air-filled and dark gray glycerol-filled parts
of the image. The image is filtered so as to obtain a clear boundary
between the two phases, through a scheme that mainly consists in
removing the background and thresholding at a gray level value between
the two latter peaks. All further image treatments are performed on the
resulting black and white image.

To check possible dynamic components of the capillary pressure we
performed gravity stabilized experiments by keeping the experimental
model vertical \cite{Meheust2002} and extracting the glycerol/water
mixture from the bottom of the model. The capillary pressure was
measured by recording the pressure in the model as the stabilized
fluid front approaches the sensor. No systematic dynamic effect on the
capillary pressure was found. For the low injection rates the width of
the fronts was further used to estimate the minimum and width $W_c$ of
the capillary pressure threshold distribution.

Throughout the paper the following coordinate system is used: ($x$,
$z'$) is the orthonormal frame describing the porous medium plane,
with $z'$ the spatial coordinate in the direction of the flow
(positive in flow direction). The position of the most advanced finger
tip is denoted $z'_{tip}$; its speed along the $z'$ axis is denoted
$v_{tip}=\dot{z}'_{tip}$. The position along the $z'$ axis computed
with respect to that of the most advanced finger tip is $z= z'_{tip} -
z'$. Those coordinates are indicated in
Fig.~\ref{fig_observations}~and~\ref{fig_subseq_subtracted}.

\section{Results}
\label{sec:results}

We present $12$ experiments using the wide model for values of the
capillary number $C_a$ ranging from $1.4\cdot 10^{-2}$ to $3.6\cdot
10^{-1}$ and $5$ experiments using the narrow model, for capillary
numbers ranging from $3.3\cdot 10^{-2}$ to $1.9\cdot 10^{-1}$. The
latter series was conducted to check system size dependencies. For
every experiment, we have carefully investigated the invasion process.

Fig.~\ref{fig_observations} displays air clusters observed for the
same porous medium, at three different flow rates. The ``fingers''
look visually thinner at increased invasion speed and more internal
trapping of the defender is observed at low speed. Hence, the
displacement exhibits obvious capillary number dependent features
which will be discussed in details in part B below. In part A, we
focus on the relation between the growth activity, the frozen
structure left behind and the pressure field in front of the fingers.

\subsection{The relation between growth activity, frozen structure and fluid pressures.}
\label{sec:screening_result}

The growth activity has been investigated by measuring the growth
probability density $\phi(z)$ from series of images and performing
pressure measurements.
\begin{figure}
 \pic{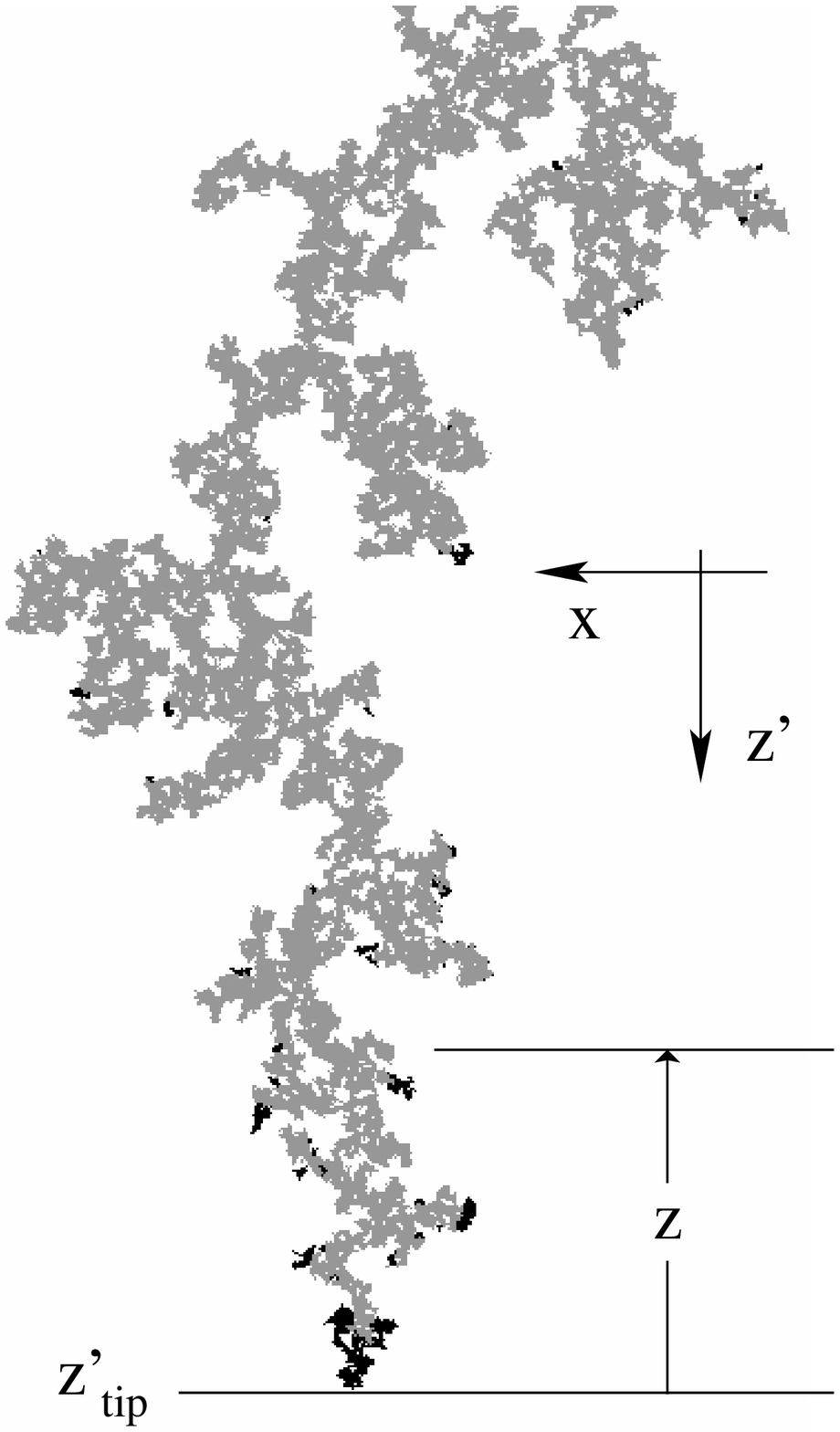}{0.6}{0}
 \figcap{fig_subseq_subtracted} { Two consecutive images taken at a
 time interval $\Delta t\simeq 15$ during the experiment at $C_a =
 0.059$, drawn on top of each other. The invaded regions in the first
 image are painted light gray, the growth areas obtained by
 subtracting the first picture from the other one are painted
 black. The coordinate system used throughout the paper is also
 shown.}
\end{figure}

\subsubsection{Definition of growth probability density $\phi(z)$ and mass density $n(z)$.}
To investigate the growth process, images have been taken with
constant time $\Delta t$ between each image. The tip position of the
longest finger is identified to find the coordinate system
$(x,z)$, and to be able to calculate the speed of the longest finger.
The differential growth between two images is found by a direct image
subtraction between two subsequent images. After the subtraction we
typically have a collection of invaded pores representing the growth
(see Fig.~\ref{fig_subseq_subtracted}) . The growth density
$\phi^*(z)$ is defined as the average number of filled pores within
$[z,z+\Delta z]$ divided with $\Delta z$. After an initial regime
corresponding to the time needed for the longest finger to propagate
the order of the width of the porous medium $W$, $\phi^*(z)$ is found
to be fairly independent of time up to a few percents variations. In a
given experiment with constant $C_a$, $\phi^{*}(z)$ is then averaged
over all images excluding this initial regime, to obtain a good
average of the stationary growth function. The growth probability
density $\phi(z)=K \: \phi^*(z) $, where $K$ is a normalization constant,
is then found by normalizing $\phi^*(z)$ with respect to $z$ such that
\begin{equation}
\int_0^L \phi(z) \: dz = 1 \;.
\end{equation}
Note that in the reminder of the paper $z$ is in units of pore
size ($a = 1$~mm). 

The mass density of the frozen structure $n(z)$ is defined as the
average number of filled pores within $[z,z+\Delta z]$ divided with
$\Delta z$. The average is taken over all images in a given experiment
with constant $C_a$. Both $n(z)$ and $\phi(z)$ appeared to be fairly
robust with respect to the width $\Delta z$ of the analysis strips
used to compute them.

\subsubsection{Growth activity and the frozen structure left behind}

For all experiments, the speed of the most advanced finger tip was
observed to be fairly constant. Fig.~\ref{fig_time_tipp} shows the
position of the most advanced finger tip $z'_{\text{tip}}$ for
different capillary numbers. After a short initial stage, the speed of
the fingers saturates to a constant average value. Linear fits to the
behavior $z'_{\text{tip}}$ as a function of time outside the 
initiation stage provide an average finger tip speed $v_{tip}$ for all
experiments.
\begin{figure}
 \pic{fig4.eps}{1}{0} \figcap{fig_time_tipp}{
 Plot of the position of the most advanced finger tip
 $z'_{\text{tip}}$ as a function of time. Data from the wide model. The
 time is rescaled by dividing with the break through time $t_b$, which
 is the time the most advanced finger reach the outlet channel. The
 values of $t_b$ are $5347$~s, $1204$~s, $476$~s and $256$~s for the
 capillary numbers $C_a$ equal to $0.027$, $0.059$, $0.12$ and $0.22$
 respectively.}
\end{figure}

\begin{figure*} 
 \bmini{0.50} (a) 
 \pic{fig5a.eps}{0.92}{0} \emini
 \bmini{0.50} (b) 
 \pic{fig5b.eps}{0.92}{0} \emini
 \figcap{fig_inv_probabl}{ Lin-log plot of the invasion probability
 density $\phi(z)$ as a function of the distance $z$ to the finger
 tip. 
 The indicated distribution corresponds to 
 {\bf (a)} data from the wide experimental model and {\bf (b)} from the
 narrow model. The solid lines in the curves corresponds to the
 model function $\ln(B e^{-z/\lambda})$ obtained from linear regression
 over $\ln(\phi(z))$ for $z/a < 100$.
}
\end{figure*}

The measured invasion probability density function $\phi(z)$ is
plotted in Fig.~\ref{fig_inv_probabl} as a function of the distance to
the finger tip for the two system sizes on a lin-log plot. An
exponential like decay is seen for $z/a<100$ with a deviation from
exponential behavior for larger lengths. A characteristic decay length
or ``screening length'' $\lambda$ is estimated from linear fits to the
lin-log data for $z/a<100$ (see Fig.~\ref{fig_inv_probabl}).
As we can see from these plots a nice data
collapse is obtained, indicating that the invasion probability density
$\phi(z)$, and thus the screening length $\lambda$, are independent of
the capillary number for a given system.  On the other hand, when
comparing the two systems, the screening length $\lambda$ depends on
the system size: $\lambda = (54\ \pm\ 10)$~mm for the wide model, and
$\lambda = (34\ \pm\ 5)$~mm for the narrow model. The actual shape of
function $\phi$ also seems to be weakly dependent on the system size.

Relating the mass of the frozen structure, $n(z)$, to the invasion
probability density, $\phi(z)$, and confronting the obtained relation
to experimental results, provides new insights into the displacement
process. The total number of invaded pores in a time interval $[t, t +
\Delta t]$ is $R\ \times \Delta t$, where $R$ is the number of invaded
pores per time unit. For a given flow rate $Q$, $R$ is related to $Q$
and to the characteristic pore volume $V_{\text{pore}}$ by the
relation $Q= R \times V_\text{pore}$, so that
\begin{equation}
 R = \frac{W}{a^2}\; v_f \text{~ ~,}
\label{eq_pore_inv_rate}
\end{equation}
where $W$ is the width of the system, $a$ is the characteristic pore
size, and $v_f$ is the Darcy or filtering velocity of the wetting
fluid; for a given porous medium and fluid pair 
$v_{\text{f}} \propto C_a$. The number of
invaded pores in the analysis strip defined by $z \in [z, z + \Delta z
]$ during time interval $[t, t + \Delta t ]$ is then $R \: \Delta t
\times \phi(z) \:\Delta z$. The tip position $z'_{\text{tip}}$ is
further given by $z'_{\text{tip}}(t) = z'_0 + v_{\text{tip}} t$ where
$v_{\text{tip}}$ is the speed of the finger tip (assumed to be
constant) and $z'_0 = z'_{\text{tip}}(t = 0)$. This is a fairly good
approximation after a short initial regime as seen in
Fig.~\ref{fig_time_tipp}.

The total number of invaded pores in an analysis strip at distance $z$
from the finger tip and in a strip of width $\Delta z$ is thus given
by:
\begin{equation}
\label{eq:int1}
 n(z) \: \Delta z = R \int_{t_0}^{t} \phi [ z(t') ] \: \Delta z \: dt'
 \text{~ ~,}
\end{equation}
where $t_0$ is the time at which $z'_{\text{tip}} = z'$. Taking
advantage of the linear relation between the coordinate $z$ and tip
speed $v_\text{tip}$, $z(t) = z'_{\text{tip}}(t) - z'_{\text{tip}}(t_0) = (t-t_0) \:
v_\text{tip}$, Eq. (\ref{eq:int1}) becomes
\begin{equation}
 n(z) =
 \frac{R}{v_{\text{tip}}} \int_{0}^{z} \phi(\tilde{z})\: d\tilde{z} 
\equiv \frac{R}{v_{\text{tip}}} \Phi(z) \text{~.}
\label{eqn:mass_dens_calc}
\end{equation}
Using Eq.~(\ref{eq_pore_inv_rate}) we finally obtain the relation
\begin{equation}
 n(z) = \frac{W v_f}{a^2\ v_{\text{tip}}}\ \Phi(z) \propto \frac{C_a}{v_{\text{tip}}} \Phi(z) \text{~~,}
\label{eq_mass_collapse}
\end{equation}
which relates the linear density of invaded pores (or ``cluster-mass''
density), $n(z)$, to the cumulative invasion probability density
distribution, $\Phi(z)$. This relation is confirmed by
Fig.~\ref{fig_nz_plot}, where $n(z) \times (a^2v_{tip}/(W v_f))$ is
plotted as a function of $z/a$. All experimental plots collapse,
confirming that there is one single cumulative probability
distribution $\Phi(z)$ for the system for all experiments at different
extraction speeds. The function $\Phi(z)$, computed as an average
function from all cumulative probability functions for the various
experiments, is plotted in Fig.~\ref{fig_nz_plot} as a plain line.

\begin{figure*} 
 \bmini{0.50} (a) \pic{fig6a.eps}{0.92}{0}
 \emini \bmini{0.50} (b)
 \pic{fig6b.eps}{0.92}{0} \emini
 \figcap{fig_nz_plot}{
 Plot of the rescaled average mass density of non wetting fluid $A
 \times n(z)$ where $A = a^2 \: v_{\text tip} / (W \: v_f)$ inside the
 model as a function of the distance to the tip of the most advanced
 finger $z$: {\bf (a)} the wide model and {\bf (b)} the narrow
 model. The average cumulative invasion probability function $\Phi(z)$
 is plotted on top of the curves for comparison in the two cases. The
 inserts of (a) and (b) show the plots $\ln \left [ 1 - ( a^2 \:
 v_{\text tip} / (W \: v_f) ) \: n(z)\right ] = \ln \left[ ( a^2
 v_{\text tip} ) / (W \: v_f) \right] + \ln \left[ (n_{\infty} -
 n(z)) \right]$ as a function of $z$. The solid lines in the inserts 
correspond to $\ln(1 - \Phi(z))$, and the dashed lines
 has the slope $1/\lambda$ where $\lambda$ is the screening length of
 $\phi(z)$ found from Fig.~\ref{fig_inv_probabl}, i.e. respectively
 $\lambda = (54
 \pm 10)$~mm and $(34 \pm 5)$ for the wide model and narrow model.
 }
\end{figure*}
The insert of Fig.~\ref{fig_nz_plot} shows $(a^2 v_{\text{tip}})/ ( W
v_f) (n_{\infty}-n(z))$ on a lin-log scale, where $n_\infty \equiv W
v_f/a^2 v_{\text{tip}}$. The solid lines represents $(1 - \Phi(z))$
and the dashed lines $(1 - e^{-z/\lambda})$ which would be the model
function for a pure exponential $\phi$ ($\lambda$ is the screening
length evaluated before).

From the results presented above we conclude that the active invasion
zone is defined by a screening length $\lambda$ which is constant for
a given porous media and liquid pair and at a range of capillary
numbers of two decades. However, we expect this result to be valid
only for sufficiently high filtration speed $v_f$. Indeed, on the one
hand, the capillary fingering regime ($C_a \simeq 0$) corresponds to an
invasion that is controlled by fluctuations in the capillary threshold
pressures, so that invasion occurs along the whole
front~\cite{Lenormand85,Lenormand89}. There is no well defined finger
tip or growth direction in that limit. The width of the capillary
threshold pressure distribution $W_c$ is larger than the viscous
pressure drop over the whole system and defining a screening length or
active zone is not meaningful.

When the length of the system is larger than its width,
 it is found from both pressure measurements and simulations
that the decay in pressure into the structure from the longest finger
occurs on a length scale of the order of the width of the system (see
Fig.~\ref{fig_observations}). We therefore expect $W$ and not $L$ to
be the relevant length scale for the decay of the pressure field close
to the tip. Viscous forces can therefore be considered to dominate
capillary pressures if the following criteria is met:
\begin{equation}
W_c < \frac{W \:\mu \: v_f}{\kappa} \text{~ ,}
\end{equation}
or if we assume $W_c \sim P_c$ (which is the case here)
\begin{equation}
 \label{eq:capn_lower_limit}
 C_a > a/W \text{~ .}
\end{equation}
For our system $a/W \sim 10^{-3}$, which is an order of magnitude
smaller than our lowest capillary number.
 
On the other hand, for situations where the ``pure viscous fingering''
in a random porous media has been reached, there is no trapping of
wetting liquid inside the fingers, which reached the lower one pore
width limit (at $C_a \approx 0.2$ in our system). Whether the
screening length or active zone have the same width or behave
identically as for lower capillary numbers is not clear. We believe
that the screening by the most advanced finger is a viscous effect,
which remains important as the displacement speed increases. In this
one pore limit however, the tip speed dependence on the capillary
number is modified, as will be further detailed in section
\ref{sec:capn_dep_result}

\subsubsection{The relation between the growth probability density $\phi(z)$ and the fluid pressure. }
\begin{figure*} 
 \bmini{0.50} 
  (a) \pic{fig7a.eps}{0.92}{0}
 \emini
 \bmini{0.50}
 (b) \pic{fig7b.eps}{0.92}{0}
 \emini
 \figcap{fig_press_sensor_data}{ Plot of the pressure difference
 $\Delta P(z)=P(z)-P(\infty)$ measured by the pressure sensor located
 in the position $(x_s,z'_s) = (38$~mm, $280$~mm$)$. The insert is the
 same data plotted in a semi-log plot where $\Delta P(z)$ is scaled
 with $\Delta P(0)=P(z=0)-P(\infty)$ to illustrate the decay of the
 pressure field. Figure {\bf (a)} is for the wide model and Figure
 {\bf (b)} is for the narrow model.}
\end{figure*}

Fig.~\ref{fig_press_sensor_data} shows the dependence of the pressure
difference $\Delta P(z) = P(z) - P(\infty)$ in the wetting liquid as a
function of the distance $z$ to the outermost tip for different
capillary numbers. Here $P(\infty)\equiv P_0 - P_c(\infty)$ is the
liquid pressure along the interface far behind the finger tip, with
$P_0$ the pressure in the non-wetting liquid and $P_c(\infty)$ the
capillary pressure in this stagnant zone. It is important to note that
the pressure $P(z)$ is measured on the side of the model (indicated in
Fig.~\ref{fig_observations}) while the fingers are
propagating in the central part of the model
(Fig.~\ref{fig_observations}).

The pressure seems to be linearly dependent on the distance from the
tip during a first stage before the tip reaches the sensor. In a
second stage, after the finger tip has passed the sensor, pressure
relaxes and reaches the value $P(\infty)$. A closer inspection of the
pressure curves (see Fig.~\ref{fig_press_sensor_data}) shows that there is
no clear systematic dependence of the pressure relaxation on the
capillary number. The pressure difference $P(z) - P(\infty)$ decays
with approximately the same length for the different capillary numbers
(see insert of Fig~\ref{fig_press_sensor_data}). This indicates that
the details of the internal structure of the ``fingers'' do not have a
strong influence on the pressure field on large scales.

The pressure measurements are related to the invasion activity by the
following considerations. Let us consider the local speed of an
interface located in an arbitrary pore throat between two pores, one
filled with air and the other with the wetting-liquid. Let $P(x,z)$ be
the pressure in the wetting liquid and $P_t(x,z)$ be the capillary
pressure threshold value to invade that pore. Note that this is
different from the pressure $P(z)$ defined as the pressure measured on
the side of the model at the sensor position. The pore throat at
position $(x,z)$ is passed under the condition that the capillary
pressure $P_{c}(x,z)=P_0-P(x,z)$ is larger than the capillary
threshold pressure $P_t(x,z)$ at this position. If invasion occurs, a
characteristic value of the speed of the interface will be:
\begin{equation}
v(x,z)=\frac{2\kappa}{\mu} \: \frac{(P_0-P(x,z)-P_t(x,z))}{a} \;.
\end{equation}
Let $N(P_t(x,z))$ be the capillary pressure distribution. For the sake
of simplicity, we assume a flat capillary pressure distribution with
lower limit $P_t^{\text{min}}$, upper limit $P_t^{\text{max}}$ and
width $W_c$. Then, the expectational value of the interface velocity
(average value over the capillary threshold distribution), while the
pore is getting invaded will be
\begin{widetext}
\begin{equation}
\left \langle v(x,z) \right \rangle = 
\frac{1}{P_0-P(x,z)-P_t^{\text{min}}} \:
\int_{P_t^{\text{min}}}^{P_0-P(x,z)} \: \frac{2\kappa}{a \mu} \: \left( 
P_0-P(x,z)-P_t(x,z) \right) \: dP_t 
= \frac{\kappa}{a\mu} \left(P_0-P(x,z)-P_t^{\text{min}}\right) \text{~ ~.}
\label{eq:local_speed_rel}
\end{equation}
\end{widetext}
Here, $P_t^{\text{min}}$ is the minimum of the distribution for
capillary threshold; when $P_0-P(x,z)$ goes to that minimum, the
expectational value for the speed of the interface goes to zero. The
growth probability density $\phi(x,z)$ for the invasion structure
within a time $[t,t+\Delta t]$ at a position $(x,z)$ is proportional
to $\langle v(x,z)\rangle$ times the probability $p(x,z)$ that the
throat gets invaded; hence
\begin{equation}
\label{invasionprobdens}
\phi(x,z)=C \left \langle v(x,z) \: \right \rangle p(x,z) \text{~ ~,}
\end{equation}
where $C$ is a normalization constant, which we can find by
integrating the above equation along the invasion front $S$:
\begin{align}
\!\!\!\!\int_S \phi(x,z) \:  dl & = C \int_S \left \langle v(x,z) \: \right \rangle p(x,z) \: dl   \\
\!\!\!\!1 & = \frac{C}{a}\int_S a \left \langle v(x,z) \: \right \rangle p(x,z) dl \; = \frac{C}{a} \: Q  
\text{~, }
\end{align}
%
where $Q$ is the flow rate, thus:
\begin{equation}
 C = \frac{a}{Q}
\label{eq:C_int_constant}
\end{equation}

Since we have assumed a flat capillary threshold distribution of width
$W_c$, the probability that the pore at position $(x,z)$ gets invaded
is
\begin{equation}
\label{probinvasion}
p(x,y)=\frac{1}{W_c} \:\left (P_0-P(x,z)-P_t^{min} \right ) \;.
\end{equation} 
From Eq. (\ref{eq:local_speed_rel} - \ref{probinvasion}) we obtain for
the growth probability density $\phi(x,z)$ in position $(x,z)$:
\begin{equation}
\phi(x,z)=\frac{\kappa}{Q \mu W_c} \: \left (P_0-P_t^{min}-P(x,z) \right )^2 \;.
\end{equation}
Averaging this expression over $x$ and introducing the number of
interface sites at a distance $z$ from the tip, $f(z)$, we obtain the
invasion probability density $\phi(z)$ as
\begin{align}
\phi(z) & = f(z) \: \left \langle \phi(x,z) \right \rangle_x \\
\label{eq:phi_press_relation}
\phi(z) & = f(z) \: \frac{\kappa}{Q \mu W_c} \left (P_0 - P_t^{\text{min}} - \left \langle P(x,z) \right \rangle_x \right )^2 ,
\end{align} 
for which we have assumed that $P(x,z)$ is a function of $z$ only
(lowest order approximation). Eq.~(\ref{eq:phi_press_relation}) yields
\begin{equation}
 \label{eq:inv_prob_press_rel}
 \left \langle P(x,z)\right \rangle_x = P_0 - P_t^{\text{min}} - \:
 \left(\frac{\phi(z)}{f(z)} \: \frac{Q W_c
 \mu}{\kappa}\right)^{\frac{1}{2}} \text{~ ~,}
\end{equation}
which can be rewritten by introducing the relation between the
flow rate and the capillary number.
Accordingly the average pressure in the wetting fluid in the immediate
vicinity of the interface and at position $z$ is related to the activity
$\phi(z)$ according to
\begin{equation}
 \label{eq:inv_prob_press_rel-2}
 \left \langle P(x,z)\right \rangle_x = P_0 - P_t^{\text{min}} - \:
 \left ( C_a \: \gamma \, W_c \: \: \frac{W}{a} \:
 \frac{\phi(z)}{f(z)} \right )^\frac{1}{2}\text{~.}
\end{equation}

Let us now look closer at the ``snapshots'' of the experiments shown
in Fig.~\ref{fig_observations}. For $z=0$, the last correction term in
Eq.~(\ref{eq:inv_prob_press_rel-2}) is $170$~Pa for the fastest
experiment ($C_a = 0.22$), and $65$~Pa for the slowest experiments
($C_a = 0.027$). At the same moment, the imposed external pressures in
the outlet channel are $3055$~Pa for the fastest and $625$~Pa for the
slowest experiments. The minimum capillary pressure is estimated to
$373$~Pa, and the width of the capillary distribution to $200$~Pa.
This indicates that the correction term in
Eq.~(\ref{eq:inv_prob_press_rel-2}) should not be neglected. In
Fig.~\ref{fig_observations} is shown the gray scale map of the
pressure field at a particular time, simulated from the displacement
structures obtained experimentally. A very strong screening is seen
for all injection rates. The large scale structure of the pressure
field in the vicinity of tip of the longest finger looks visually very
similar even if the invader structure is quite different. In the
simulations of the pressure field we have used
Eq.~(\ref{eq:inv_prob_press_rel-2}) to set the proper boundary
conditions. The pressure field has been calculated by solving the
Laplace equation for the pressure using a conjugate gradient method
\cite{Batrouni88}. We used the boundary condition given by
Eq.~(\ref{eq:inv_prob_press_rel-2}) on the cluster and the inlet line.
As boundary condition on the outlet we used the pressure
$P(\infty)-\Delta P_{\text{tot}}$, where $-\Delta P_{\text{tot}}$ is
the total viscous pressure drop imposed in the corresponding
experiment at that moment. To obtain $P(\infty)=P_0 - P_c(\infty)$,
the capillary pressure $P_c(\infty)$ was measured in the experiments
for large values of $z$. Fig.~\ref{SimPressure} shows the simulated
pressure $\Delta P(x_s,z) = P(x_s,z) - P(\infty)$ as a function of the $z$--coordinate
 relative to the tip position defined as previously, at a fixed lateral position
 $x_s$ corresponding to the $x$--coordinate of the pressure sensor.
 It is important to note that this is somewhat different
from the experiments since the pressure is measured at different $z'$
positions, but at the same time, i.e. with a fixed geometry of the invasion cluster,
 while in the experiments the pressure is measured at a fixed $z'$ position, 
at different times corresponding to various stages of the invasion cluster.
 The length scale of the decay of the pressure for $z>0$ is very similar in the experiments
and the simulations (see comment below). However the pressure difference $\Delta P(z)$ in the
simulations is lower than the $\Delta P(z)$ measured by the sensor (at
position $(x_s,z'_s)=(38$~mm, $280$~mm$)$) in the model (see
Fig.~\ref{fig_press_sensor_data}). This is due to the strong boundary
effects of the pressure close to the outlet channel: the tips in the simulations situations 
are very close to this boundary along which the pressure is fixed (see Fig.~\ref{fig_observations}),
 while in the  situations corresponding to the measurements at $z<0$ plotted 
in Fig.~\ref{fig_press_sensor_data}, 
the outlet boundary was far ahead of the finger tip, and the pressure boundary condition
 was equivalent to an imposed gradient at infinite distance.
To check the importance of this boundary effect on the magnitude of the pressure difference,
 the simulated pressure has been compared with the
pressure difference $\Delta P(z)$ evaluated from measurements at the outlet channel,
 as the finger tip progressed further than the stage corresponding to the simulations. 
The agreement in Fig.~\ref{SimPressure} between  the simulated pressure 
 and the data points corresponding to the outlet channel measurements is then satisfactory.

\begin{figure}
 \pic{fig8.eps}{1}{0}
 \figcap{SimPressure}{ The simulated pressure along the line with
 $x_s=38$~mm (corresponding to the $x$ coordinate of the pressure
 sensor) for the invasion structures in
 Fig.~\ref{fig_observations}. The data points in the main graph
 show the corresponding pressures measured at the outlet sensor.
 In the insert we plot scaled simulation data with corresponding
 pressure data measured inside the model at 
 $(x_s, z'_s) = (38$~mm, $280$~mm$)$.}
\end{figure}

To compare the length scale of the decay of the pressure for $z>0$ between the simulations
and the experiment, we then compare the pressure data measured inside the model at sensor position,
 to the simulation data scaled by a factor such as $\Delta P(x_s,z=0)$ would be equal in experiments and simulations.
Such rescaling of the simulation pressure profile simply corresponds to the result of an identical simulation
still carried on the invasion clusters of Fig.~\ref{fig_observations}, with identical boundary conditions  derived from the growth density function for the pressure along 
the clusters, but where the imposed pressure along the bottom boundary is such that the pressure at point
$(x_s,z'_{tip})$ would coincide with the pressure measured in the experiments when the tip passed at the same height as the sensor, i.e. when $z'_{tip}=z'_s$. 
This ensures that the pressure gradient and pressure value in the region around the tip of the invading cluster are of the same order in these rescaled simulations and in the
experimental stages corresponding to $z \sim 0$ in Fig.~\ref{fig_press_sensor_data},  which is a first order technique to correct for the strong boundary effect and compare with these experimental situations where the bottom boundary is much further away.
 The pressure measured in the experiments at sensor position and this scaled simulation data are plotted in the insert of Fig.~\ref{SimPressure}.
This comparison shows that the decay in the pressure happens at comparable length scales in the simulations and experiments.

 Eventually, the local structure of the finger and the lateral $x$--distance from the invader to the pressure sensor
will also have an important influence on the pressure field. The
difference in the pressure field between the left and the right side
of the finger (looking in the flow direction) in
Fig.~\ref{fig_observations} illustrates this point. The deviation
between the experimental data points and the simulations for the
lowest capillary number of the main part of Fig.~\ref{SimPressure} may be explained by
this effect. As the lateral position $x_{tip}$ of the invading structure moves
during the experiment, and is importantly varying from an experiment
to the next, this effect also explains the important dispersion of the
scaled pressure drops $\Delta P(z) / \Delta P(0)$ observed in the
insert of Fig.~\ref{fig_press_sensor_data} (b).

\subsection{Capillary number dependent features}
\label{sec:capn_dep_result}

As stated in the introduction to section \ref{sec:results},
Fig.~\ref{fig_observations} clearly shows that some features of the
invading cluster depend on the capillary number. The mass $n(z)$
of the invasion cluster obviously decreases with increasing capillary
number; in relation to this, the speed of the most advanced finger
tip, $v_\text{tip}$, increases with the capillary number, and there is
a systematic trend for fingers to become thinner as capillary number
increases. In the following we first present results relative to the
``mass density'' in the stagnant zone, $n_{\infty}$, and to the
velocity of the most advanced fingertip, $v_\text{tip}$. In the end we
discuss the results relative to measurements of the characteristic
width of the finger-like structures, the definition of which are not
as straightforward and clear as those of $n_{\infty}$ and
$v_\text{tip}$.

%
%
\begin{figure*}
 \bmini{0.49} (a) 
 \pic{fig9a.eps}{0.9}{0} \emini \bmini{0.49} (b)
 \pic{fig9b.eps}{0.9}{0} \emini
 \figcap{fig_nst_vtip_scaling}{ Log-log plots of (a) the saturated
 mass density, $n_{\infty}$, and (b) of the speed of the most advanced
 finger , $v_{\text{tip}}$, as a function of the capillary number, for
 the two set of experiments. Both plots are consistent with a scaling
 in the form $n_{\infty} \propto C_a/v_{\text{tip}} \propto
 C_a^{-\alpha}$, with $\alpha = 0.65 \pm 0.05$.}
\end{figure*}
The evolution of the average mass density in the stagnant zone
$n_{\infty}$ as a function of the capillary number is presented in
Fig.~\ref{fig_nst_vtip_scaling}~(a) on a log-log scale. The data is
consistent with a scaling law in the form $n_{\infty} \propto
C_a^{-\alpha}$, with a scaling exponent $\alpha = 0.65 \pm 0.05$ for
both the wide and narrow models. 
Here $n_{\infty}$ has been measured by fitting the function
$n_{\infty} \:\left[1 - \exp(-z /\lambda) \right]$ with both
parameters free to our measured $n(z)$ data. Due to that dependence of
the mass of the invasion cluster on the capillary number
(Fig.~\ref{fig_nst_vtip_scaling}~(a)), the speed of the most advanced
finger tip, $v_{\text{tip}}$, is expected to depend on the filtration
speed or capillary number in a non-linear way. The saturated mass
density and the speed of the most advanced finger are related to each
other through Eq.~\ref{eq_mass_collapse}, according to
\begin{equation}
 v_{\text{tip}} \propto \frac{C_a}{n_{\infty}} \text{~ ~.}
\label{eq:vtip_ninfty_rel}
\end{equation}
Based on that argument, 
$C_a/v_{\text{tip}}$ should therefore scale in the same way as
$n_{\infty}$ with respect to the capillary number. In
Fig.~\ref{fig_nst_vtip_scaling}~(b), the quantity $C_a/v_{\text{tip}}$
is plotted as a function of $C_a$ using a log-log scale. The plot is
consistent with the expected scaling (\ref{eq:vtip_ninfty_rel}) and
the result for the mass density presented above.

\begin{figure}
 \pic{fig10.eps}{1}{0} \figcap{fig_wf_scaling}{
 Log-log plot of the measured characteristic width of the finger like
 structures as a function of the capillary number, for both the wide
 and narrow model. The data is consistent with a scaling of the finger
 width in the form $w_f \propto C_a^{-\beta}$, with $\beta = 0.75 \pm
 0.05$. }
\end{figure}

The study of the dependency of the finger width on the capillary
number is somewhat less straightforward, because our invading clusters
structures exhibit extensive branching and display ``fingers'' both at
small scales as ``capillary fingers'' and at large scales as ``viscous
fingers''. Thus, a precise definition of a finger, and furthermore a
finger width, is not an easy task for those structures. A possible
method to determine the viscous finger width would consist in finding
the characteristic crossover length between geometric features
characteristic of viscous fingering and those characteristic of
capillary fingering from the density-density correlation function of
the structures. However, due to the small difference in fractal
dimension between the two regimes, $1.83 \pm
0.01$~\cite{Lenormand85,Lenormand89} for capillary fingering and $1.62
\pm 0.04$~\cite{Maloy85,Chen85} for viscous fingering, larger systems
would be necessary for this method to be accurate enough.
An experimental determination of the characteristic width $w_f$ for
viscous fingers was previously obtained for {\em imbibition}
experiments \cite{Stokes86}, for which the characteristic finger width
can be defined and found in a more straightforward manner. The
obtained scaling was $w_f \propto C_a^{-0.5}$. In those imbibition
experiments, the finger width $w_f$ was measured as the average length of
cut--segments perpendicular to the flow direction.
This method can also be applied in our experiments, but due to the small
scale fractal nature of the invasion front, trapping of wetting fluid
inside the fingers and capillary fingering on small length scales, it
is not obvious which length scales are being probed with this method.
The results that we obtain are plotted as a function of the capillary
number in Fig.~\ref{fig_wf_scaling}. Clusters of wetting liquid
trapped behind the displacement front have been removed from the
picture prior to analysis. We then define the front width $w_f$ as the
average over $z$ and time of the length of the intersects 
between the invasion cluster emptied from these trapped regions and cuts perpendicular to the flow direction.

The measurements
are consistent with a scaling law in the form $w_f \propto
C_a^{-\beta}$, with $\beta = 0.75 \pm 0.05$. This is significantly
different from what was measured for imbibition. It also differs
significantly from the scaling law expected from theoretical arguments
for percolation in a destabilizing
gradient~\cite{Lenormand89b,Yortsos2001} for two-dimensional systems:
$w_f \propto C_a^{-\beta}$ with $\beta = 0.57$. In our experiments,
the destabilizing field (pressure) is highly inhomogeneous, which may
explains why the behavior expected from the percolation in a gradient
theory is not really observed.

From Fig.~\ref{fig_nst_vtip_scaling}~and~\ref{fig_wf_scaling}, the
observed scalings appear to be valid for a limited range of capillary
numbers.
For high capillary numbers the observed scaling breaks down for
$C_a \approx 0.2$, which corresponds to situations where the characteristic
finger width have reached the one pore limit.
At the other limit, for small capillary numbers we don't
seem to reach the lower limit in capillary number. But we expect that
the observed scaling brake down for capillary numbers smaller than the
criteria given in
Eq.~(\ref{eq:capn_lower_limit}), $C_a \sim 10^{-3}$.

\section{Conclusion and Prospects}
\label{sec:discussion}

We have studied the dynamics of the invasion process observed during
drainage in a two-dimensional porous medium, for extraction speeds
that result in an unstable fingering of the displacing non-wetting
fluid into the displaced wetting fluid.

Our main finding is that for a given porous medium, the displacement
is controlled by an invasion probability density 
that only depends on the distance of the point where it is measured to the
tip of the most advanced finger tip, and is independent of the
capillary number. The decay of this invasion probability density,
$\phi(z)$, defines an active zone for the invasion process, outside of
which the viscous pressure field can be considered to be screened by
the invasion structure. In particular, parts of the invasion structure
lying outside this active zone are frozen and do not evolve in time
any more. The size of the active zone, of characteristic screening
length, $\lambda$, was found to be independent of the capillary number
for a wide range of injection rates. In addition, experiments carried
out on models with two different widths suggested that the invasion
probability density appears to be capillary number independent, its
actual shape being possibly fixed by the system size. While the invasion
process is described by an invasion probability density that is
independent of the capillary number, the invasion speed and displaced
volume in the stagnant zone were found to scale on the capillary
according to power laws, $n_{\infty} \propto C_a/v_{\text{tip}}
\propto C_a^{-0.65}$.

The link between the growth probability and the pressure field has 
been studied. An expression for the pressure boundary condition on the
cluster has been calculated which relates the pressure on the
interface of the invader to the growth probability density function
$\phi(z)$. The measured pressure has been compared to the
corresponding simulated pressure by solving the Laplace equation for
the pressure field using this expression for the boundary condition on
the cluster. A good agreement is found between the simulations and the
experiments.

System size dependencies should be subject to further investigations,
both experimentally and by means of computer simulations.

%
%
\section{Acknowledgments}
\label{sec:acknowledgments}
The work was supported by NFR, the Norwegian Research Council, VISTA,
the Norwegian academy of science and letters' research program with
Statoil and the French/Norwegian collaboration PICS.

\bibliographystyle{apsrev}


\end{document}